\documentclass[english,superscriptaddress,twocolumn,PRE]{revtex4-1}
\usepackage[latin9]{inputenc}
\usepackage{babel}
\usepackage{docs}
\usepackage{bm}
\usepackage[colorlinks=true,linkcolor=blue]{hyperref}
\usepackage{color}
\usepackage[usenames,dvipsnames]{xcolor}
\usepackage{graphicx}
\usepackage{dcolumn}
\usepackage{natbib}
\usepackage{subfigure}
\usepackage{relsize}
\usepackage{amssymb,amsmath}
\usepackage{epstopdf}
\usepackage{wrapfig}
\usepackage{float}
\usepackage{yfonts}
\usepackage{todonotes}
\usepackage{setspace}
\usepackage{latexsym}

\usepackage{graphicx,color}



\begin{document}

\title[Microorganism billiards in closed plane curves]{Microorganism billiards in closed plane curves}
\thanks{Special thanks to Yuliy Baryshnikov, Sergei Tabachnikov, Vadim Zharnitsky, Maxim Arnold and Saverio Spagnolie for helpful comments and guidance.}
%
\author{Madison S. Krieger}
\email{madison$_$krieger@brown.edu}
\affiliation{School of Engineering, Brown University, Providence, RI 02912 USA}

\date{\today}
%
\begin{abstract}
Recent experiments have shown that many species of microorganisms leave a solid surface at a fixed angle determined by steric interactions and near-field hydrodynamics. This angle is completely independent of the incoming angle. For several collisions in a closed body this determines a unique type of billiard system, an aspecular billiard in which the outgoing angle is fixed for all collisions. We analyze such a system using numerical simulation of this billiard for varying tables and outgoing angles, and also utilize the theory of one-dimensional maps and wavefront dynamics. When applicable we cite results from and compare our system to similar billiard systems in the literature. We focus on examples from three broad classes: the ellipse, the Bunimovich billiards, and the Sinai billiards. The effect of a noisy outgoing angle is also discussed. 
\end{abstract}
%
%
\maketitle

\section{\label{sec:microint}Introduction}

Chaotic billiards~\cite{chernmakar,wojt1986,donnay1991,tabachbook} continue to be an extremely active area of research and its objects of study are of interest in varied disciplines of mathematics. Billiard systems are also useful tools and models in physics~\cite{chatterjee96,bhpz2004,ulambilliard}. Current extensions to well-studied billiard problems include modifications of the table geometry, the shape of the inter-collision trajectories, and the rule for generating a new trajectory upon contact with the table boundary. In this direction, recent attention has been paid to aspecular reflection laws, especially in dissipative billiard systems commonly referred to as pinball billiards~\cite{ams2009,mps2009,ams2012,mddgp2012} and slap maps~\cite{mddg2015,mddgp2014,mddg2014}, as well as to aspecular reflection laws arising from other physical effects~\cite{adh2008}. 

Recent experiments have revealed that microorganisms also play mathematical billiards~\cite{kdpg2012}. This was a surprising result in the field of microorganism locomotion and fluid dynamics; the role of boundaries has been known for some time to have a strong effect on microorganism trajectories, even at large distances --- results on mammalian spermatozoa~\cite{rothschild1963,fm1995,smithblake2009,sgbk2009} as well as on bacteria and self-propelled particles~\cite{lpcp2010,mbss2014,lauga06,ElgetiGompper2009,GiaccheIshikawaYamaguchi2010,hdf1992,btbl2008,ShumGaffneySmith2010} gave rise to a standard theory on the reorientation of microorganism trajectories in the presence of solid boundaries. The theory~\cite{LaugaPowers2009,sl2012} can be summarized succinctly: far-field hydrodynamics cause microorganisms to reorient themselves either parallel or perpendicular/antiperpendicular to a solid boundary, depending on the sign of their hydrodynamic dipole. This is the leading-order term of the Green's function for the appropriate hydrodynamic equation describing the global flow caused by the shape-deforming microorganism. The two possibilities are commonly referred to in the literature as being either of ``pusher'' type, describing microorganisms which experience a drag force on the head (or foremost body part in the direction of motion) and propel from the rear, or of ``puller'' type, which propel from the head and experience the largest drag on the rear. The theory predicts that pushers parallelize with solid boundaries; some common microorganisms of this class include spermatozoa and \textit{E. coli}, while pullers orient along the normal line to the solid boundary; some model organisms from this class include many algal cells such as \textit{Chlamydomonas reinhardtii}, which locomotes by performing a breaststroke-type motion with two fore flagellar appendages. 

While this theory is accurate for many microorganisms (especially prokaryotes), it seems that scattering from solid boundaries is completely different for the microorganisms studied in the experiment (mostly eukaryotes, whose flagellar structure is much more complex). For these organisms, the scattering events are defined by near-field hydrodynamics and steric interactions; the swimmer collides with the solid boundary and is reoriented a certain amount depending on its morphology, which includes both the body geometry (species) and the flagellar/ciliar expression (which depends on temperature, possible mutations, and so forth). During this process, memory of the incoming angle is lost; after the collision, the outgoing angles are governed by a distribution with well-defined peaks. This is the motivation for the current work, which considers a model of such collision events as an aspecular reflection law for two-dimensional closed billiard tables with a single moving particle (the microorganism). This extends previous work~\cite{wlst2016} on a similar billiard system in some two-dimensional polygonal tables. The main results of that work which differ from classical polygonal billiards were of a hyperbolic nature; for different tables and outgoing angles, regions of large negative Lyapunov exponent with periodic attractors were seen, and some very small regions of positive Lyapunov exponent correlated with ergodic components (unlike non-hyperbolic ergodic orbits in polygonal tables with specular reflection). Because the aspecular law prohibits us from using several quantitative tools, especially those involving invariant measures, we focus primarily on numerical simulations. When configuration spaces are plotted, our code for generating these graphs is based on previous work~\cite{lanselcode}. Our work may also be applicable to the design of robots which re-orient themselves based on tactile rather than visual data. These two cases of aspecular billiards have developed in parallel; for instance, the motion of aspecularly-reflecting robots~\cite{el2013} is analogous in certain respects to microorganism billiards, while rectification of ensembles of such locomoters by gears have been considered in both in the microorganism case~\cite{adr2009,saga2010,ladriscmaf2010} and in the robotic case~\cite{lz2013}.

The organization of the paper is as follows: in Sec. \ref{sec:theory}, we define the microorganism billiard and assess the quantities of interest. The aspecular nature of the reflection law prohibits the use of several analytical tools known in the literature, so these quantities are primarily determined via numerical simulation, though we do provide qualitative results from the theory when possible. We then provide three examples from billiards classes of interest: the ellipse (Sec. \ref{sec:ellipse}), the Bunimovich stadium (Sec. \ref{sec:stadium}), and the Sinai billiard in a square unit cell (Sec \ref{sec:sinai}). We summarize the results in Sec. \ref{sec:summary}.


\section{\label{sec:theory}Microorganism billiards}
\begin{figure}
\centering
\includegraphics[width=0.3\textwidth]{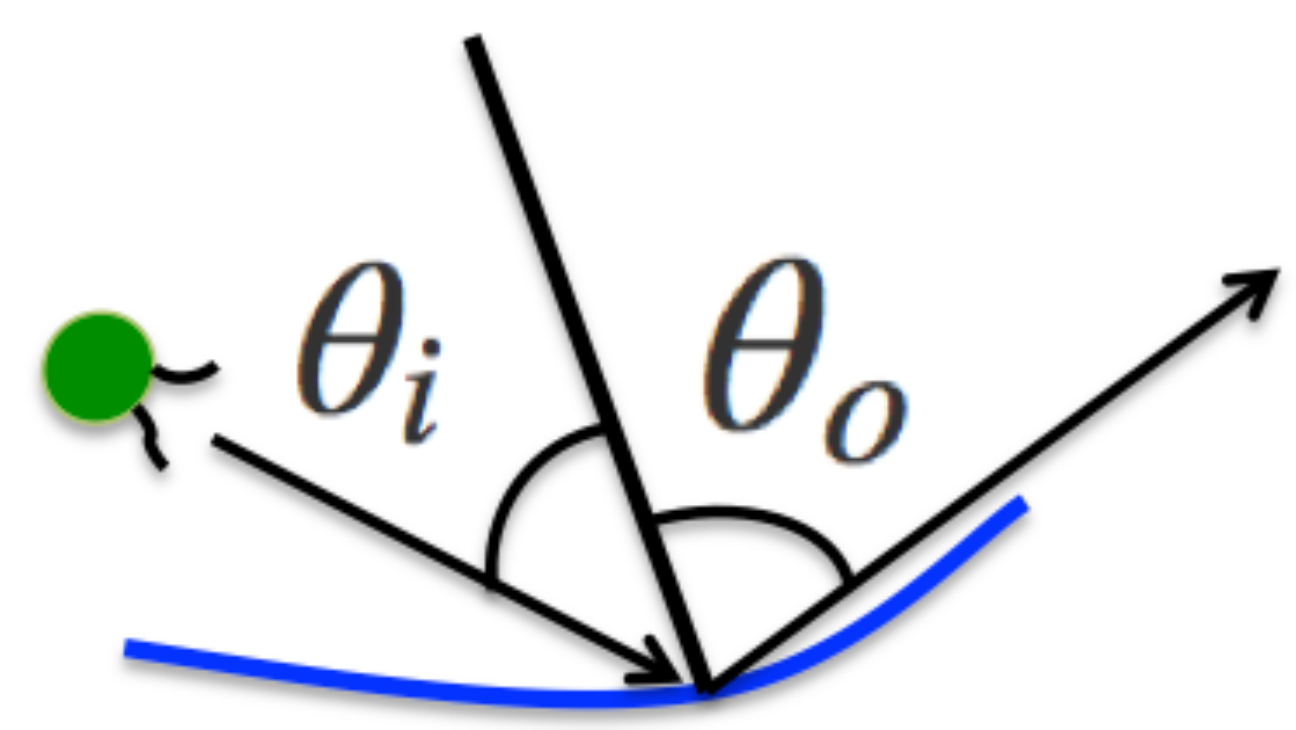}
\caption{(Color online) Geometry of microorganism billiards. The microorganism collides with the curved boundary of the table at an angle $\theta_i$ and departs at a fixed angle $\theta_o$. All angles are measured with respect to the normal at the point of collision. }
\label{setupfig}
\end{figure} 

We consider two-dimensional billiard tables populated by a single particle which moves with constant (unit) speed and whose free-flight path is always a straight line. The only departure from a classical billiard is our reflection law, which is aspecular: rather than the specular reflection law $\theta_o=\theta_i$, where alphabetical subscripts $\{o,i\}$ refer to ``out'' and ``in'', respectively, our reflection law is $\theta_o=C$, where $C$ is a constant or a random variable with a prescribed distribution. Because the outgoing angle is fixed, the microorganism billiard is effectively a one-dimensional map~\cite{demelostrein}. However, unlike the case of bouncing robots~\cite{el2013}, we choose our reflection law to preserve orientation of the trajectory with respect to the tangent line immediately before and after a collision; that is, the microorganism always crosses the normal\footnote{It is true that in the aforementioned experiments this condition is not always satisfied, however such events seem to be somewhat rare and confined to particular species, so we ignore them in this initial work}. The outgoing angle is therefore one of two possibilities, because crossing the normal can change the orientation of the following links from clockwise to counter-clockwise depending on the incoming angle and the table curvature. Equivalently, the map is only completely determined at a boundary point $q$ if the pre-image of that point is also given, as these two pieces of data determine the subsequent orientation. The one-dimensional map is therefore only a map of the true collision space when this orientation is preserved for all collisions. We point out the ramifications of these orientation reversals and when the dynamics is completely or only partially encapsulated in the appropriate one-dimensional map $F(q)$, where $q$ is the arclength parameter for the table boundary. The derivative of this map is given~\cite{ams2009,mps2009} by 
\begin{equation}
F'(q_0)=-\frac{t K_0 + \cos \theta_{o,0}}{\cos \theta_{i,1}}, \label{onedimmap}
\end{equation}
where the numeric subscripts $\{0,1\}$ indicate that the map $F:q_0 \rightarrow q_1$ takes a position on the table boundary $q_0$ with boundary curvature $K_0$ and (fixed) outgoing angle $\theta_{o,0}$ with respect to the normal to the table boundary at $q_0$ and maps it to a new boundary position $q_1$. When a straight line is drawn in the table between $q_0$ and $q_1$, representing the actual billiard trajectory, it has length $t$ and intersects the table boundary at an angle $\theta_{i,1}$ with respect to the normal to the boundary at $q_1$. To avoid a cumbersome amount of indices, we write $\eta \equiv \cos \theta_{i,1}$, and henceforth the numeric subscripts of $0$ on all other quantities will not be written. 

The unusual reflection law has strong implications for the toolbox which is available to analyze microorganism billiards. Perhaps the most important is that these systems do not preserve area in phase space, and indeed have no natural invariant measures. As a consequence, we do not speak at all to possible ergodic and statistical properties of microorganism billiards, since it is not clear how to derive results in these directions in the absence of such a measure. In addition, the involutive or time-reversal property of specular billiards is not preserved in microorganism billiards, and thus we must be extremely cautious in invoking pre-images or continued fractions reaching backwards in time from the current point in a microorganism's orbit. This also restricts the extent to which we can truly utilize one-dimensional dynamics; we discuss this on a case-by-case basis in the text.

Unlike previous work on microorganism billiards~\cite{wlst2016}, we do not consider the role of ``skids'', or events in which the billiard particle collides with the wall at an arclength position $q$ and departs at an arclength position $q \pm s$, where $s$ is the length of the ``skid''; one of the conclusions of that work was that skids had negligible impact on the dynamics. In our system, it would be possible to ``skid'' from regions of low curvature to regions of high curvature in some tables; however, recent experiments~\cite{szs2015} suggest that surface roughness reduces the length of the skid dramatically, so we consider our billiard tables to have roughness on a small scale. This assumption may affect the outgoing angle of a true microorganism; we assume that re-orientations due to the roughness of the table are incorporated as noise into the distribution of the outgoing angle. The complete problem is schematized in Fig. \ref{setupfig}. 

We will focus our efforts on quantities that we believe will be of interest to both the dynamics and the physics communities. We compare the orbit structures in various tables to the classical billiard problem to understand the potential behavior of a microorganism as well as to extract visual information about the topological dynamics, such as rotation numbers of periodic orbits. We use this information to construct a phase diagram of the different behaviors as a function of table geometry and outgoing angle, and also use this information to design useful microfluidic sorting and trapping mechanisms that might find application in experiment. Motivated by results~\cite{wlst2016} suggesting that the hyperbolic dynamics of a microorganism billiard differs dramatically from a classical billiard, we place special emphasis on Lyapunov exponents. These will be calculated using standard methods~\cite{oseledets} from specular billiards simulation and theory rather than the one-dimensional map, due to the possibility of orientation-reversals occuring multiple times over many collisions. 


\section{\label{sec:ellipse}Ellipse}
\subsection{\label{subsec:ellone}One-dimensional dynamics}

\begin{figure}
\centering
\includegraphics[width=0.5\textwidth]{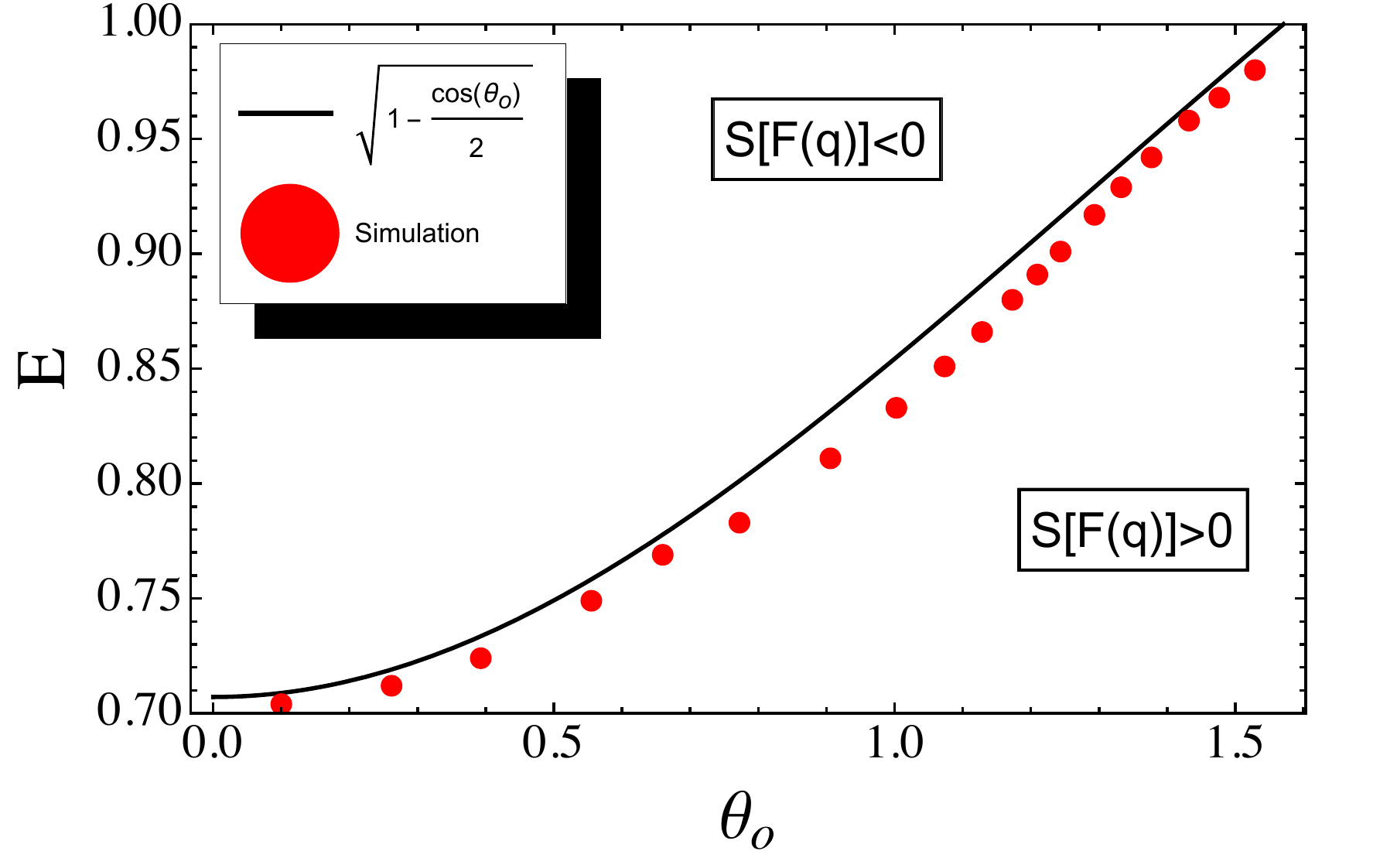}
\caption{(Color online) Diagram of the regions of negative and positive Schwarzian derivative for the one-dimensional map. The solid line indicates the curve given by Eq. (\ref{critellipse}), above which the numerator of Eq. (\ref{onedimmap}) cannot change signs. The red dots indicate the lowest value of $E$ for a given $\theta_o$ where a globally piecewise-negative Schwarzian derivative is seen in the numerical simulations. }
\label{schwarz}
\end{figure} 


\begin{figure}[h]
\centering
\includegraphics[width=0.5\textwidth]{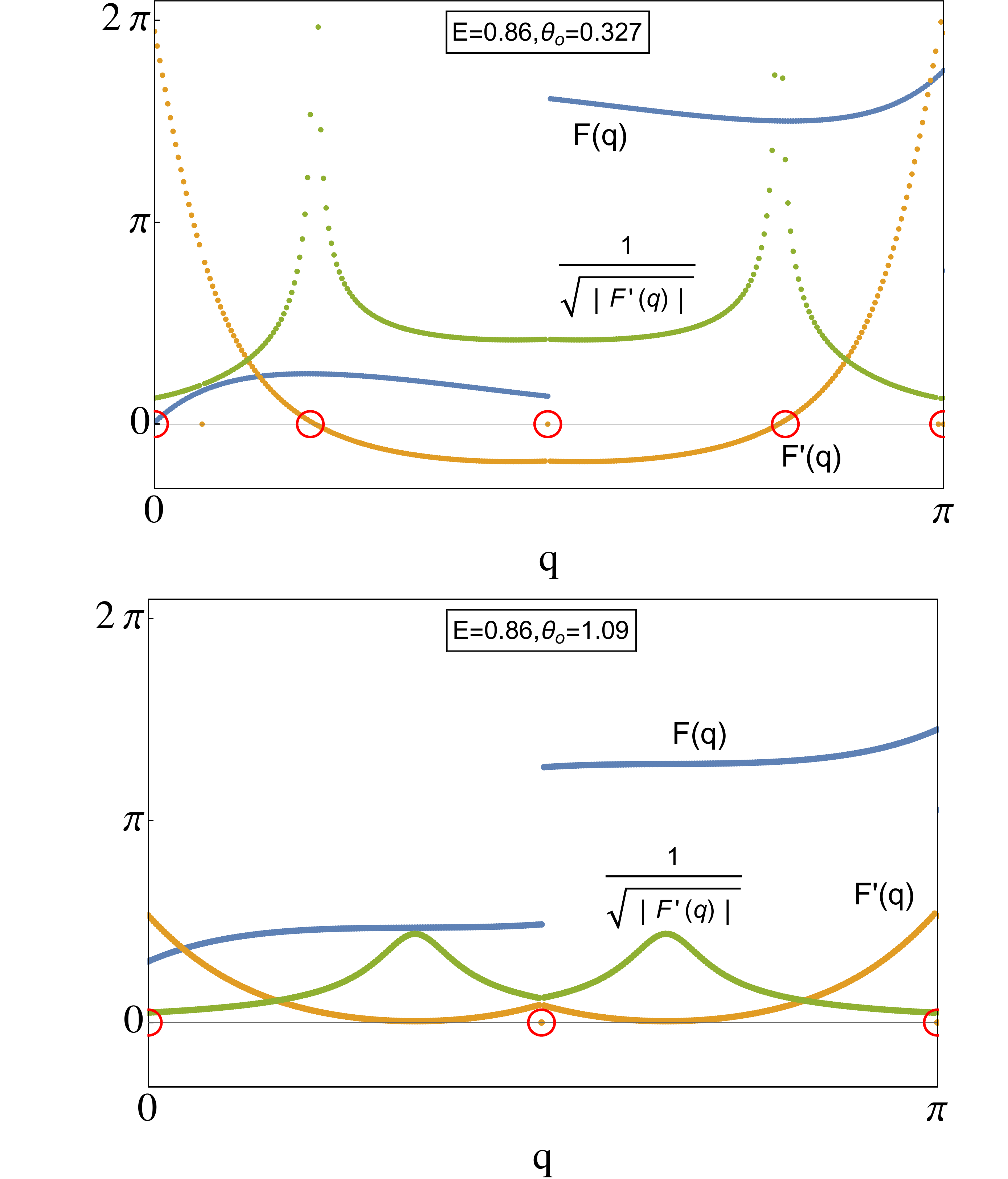}
\caption{(Color online) Plots of the billiard map $F(q)$ (blue), its deriviative $F'(q)$ (orange), and the quantity $\frac{1}{\sqrt{F'(q)}}$ (green), which is convex when the Schwarzian derivative is negative. The plot is shown on the range $q \in [0, \pi]$ and is extended to the full table by symmetry. $Sing(F)$ is denoted with red circles. The plots are shown in an ellipse with $E=0.86$ and $\theta_o=0.327$ (top), $\theta_o=1.09$ (bottom). These values have been chosen such that the top panel displays a pair $(E,\theta_o)$ lying above the curve Eq. (\ref{critellipse}), while the lower panel displays a pair below this transition point. This illustrates the global transition from a one-dimensional billiard map with piecewise negative Schwarzian derivative to one with piecewise positive Schwarzian derivative. In crossing the curve shown in Fig. \ref{schwarz}, all critical points vanish, changing the behavior of the Schwarzian derivative. }
\label{ellipsetriptych}
\end{figure} 

We begin with an analysis of the one-dimensional map, Eq. (\ref{onedimmap}), specialized to the case of the ellipse. The map is only piecewise continuous; we denote the singular set, comprised of critical points and points of discontinuity, by $Sing(F)$. An example of such a set is shown in Fig. \ref{ellipsetriptych}. We are interested to know the regime in which the Schwarzian derivative of this map is negative. This simplifies the dynamics dramatically, for negativity of the Schwarzian is preserved under forward iterations of the map and much is known in the literature~\cite{demelostrein,bpp2016} about hyperbolicity and the structure of attractors for such maps. The reader can easily check that the Schwarzian derivative $S(F(q))=\frac{F'''(q)}{F'(q)}-\frac{3}{2}\left(\frac{F''(q)}{F'(q)}\right)^2$ is negative if the quantity $\frac{1}{\sqrt{|F'(q)|}}$ is convex; thus, this behavior can be deduced by analysis of Eq. (\ref{onedimmap}). After lengthy algebra it is possible to demonstrate that the only way for $F'(q)$ to preserve one sign for all values of $q$ is if $\mathrm{min}(t K) > \cos \theta_o$. In terms of the eccentricity~\footnote{\label{note1} For all tables under consideration we assume the speed of the particle scales with the table size, so that the table geometry can be specified by a single nondimensional number} $E = \sqrt{1-b^2/a^2}$, where $a$ is the length of the major axis and $b$ is the length of the minor axis, the critical ellipse separating the two sign behaviors of $F'(q)$ has eccentricity
\begin{equation}
E_c = \sqrt{1-\frac{\cos \theta_o}{2}}. \label{critellipse}
\end{equation}
By sampling a few values of $E$ and $\theta_o$ above this line, it is easy to see that the quantity $1/\sqrt{|F'(q)|}$ is always convex away from the singular points, where the Schwarzian derivative diverges; thus all microorganism billiards above this curve have negative Schwarzian derivative $\forall q \notin Sing(F)$. 

To confirm this, we integrated Eq. (\ref{onedimmap}) numerically to obtain $F(q)$. Note that due to the symmetry of the ellipse, we can describe orientation-preserving links using iterations of $F(q)$, and can include orientation reversals by using reflections across the axes. A generic orbit is therefore described by a word on two generators of the type $FFFRFRFFR...$, where $F$ is the billiard map and $R$ is a suitable reflection. We measured the Schwarzian derivative of $F$ and took note of the first value of $E$ for a fixed $\theta_o$ when the Schwarzian derivative became globally negative. The values collapse well on to the curve predicted by Eq. (\ref{critellipse}), see Fig. \ref{schwarz}. Below this curve, the Schwarzian derivative is positive $\forall q \notin Sing(F)$ --- surprisingly, the Schwarzian has only one sign $\forall q \notin Sing(F)$ for every ellipse and outgoing angle away from the critical line. Note that the pre-image of any point under $F$ is unique in the ellipse, so that the inverse is well-defined; this means that the curve Eq. (\ref{critellipse}) also defines the sign of the Schwarzian derivative for the inverse map, which is the opposite as the sign for the forward map. However, for the actual microorganism billiard, information about the inverse map is only useful if there are no changes in orientation. To provide an illustration of how Eq. (\ref{critellipse}) organizes the (piecewise) global sign of the Schwarzian derivative, Fig. \ref{ellipsetriptych} shows the maps $F(q)$, $F'(q)$, and $\frac{1}{\sqrt{|F'(q)|}}$ for one ellipse and three different values of $\theta_o$ which lie on different sides of the critical curve. The disappearance of several points from $Sing(F)$ is responsible for the global change in sign. 

\begin{figure}[h]
\includegraphics[width=0.5\textwidth]{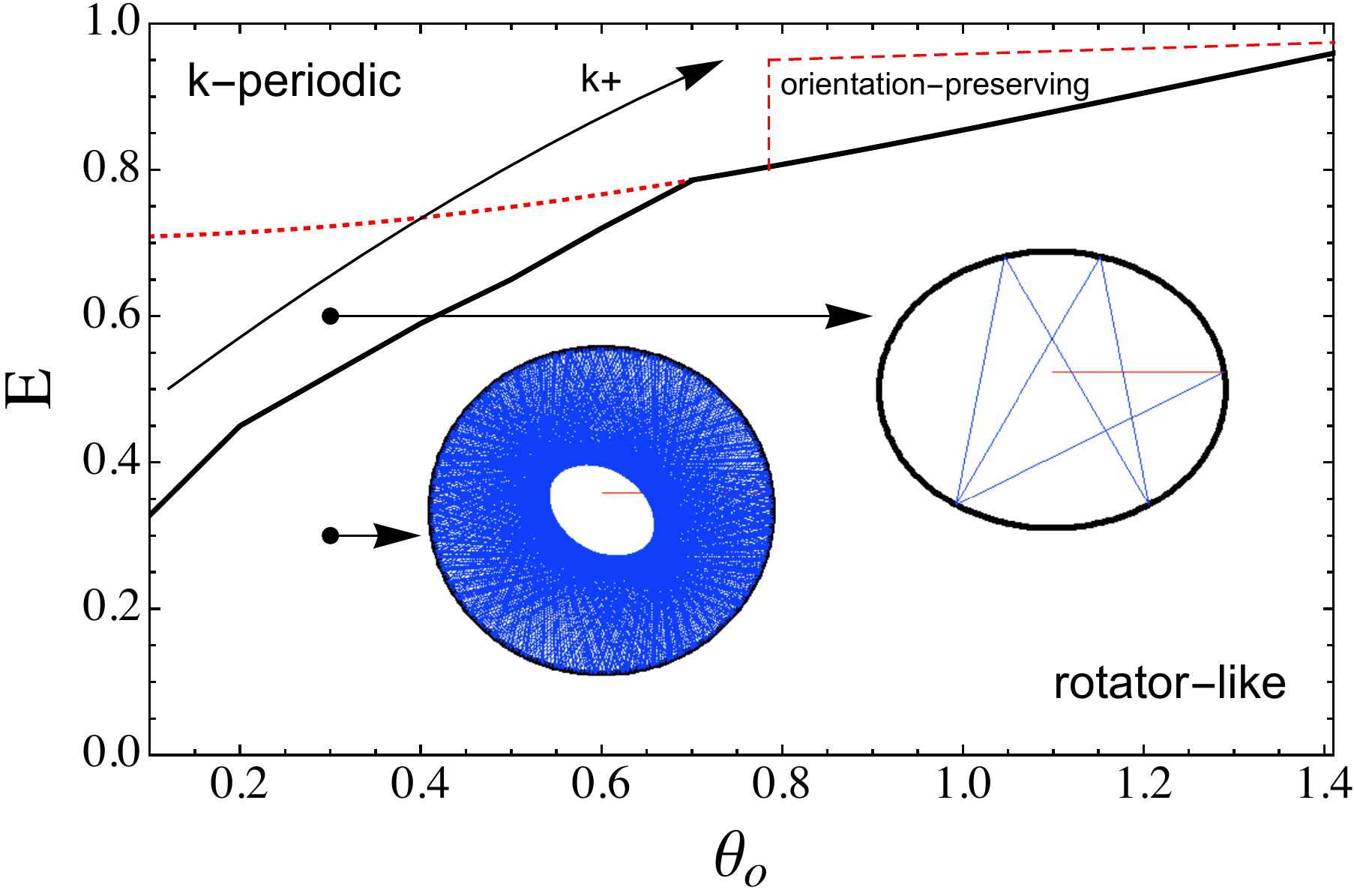}
\caption{(Color online) The phase diagram for microorganism billiards in the ellipse. The solid black line indicates the division between periodic orbits and rotator-like orbits whose limit set is dense. This line is identical to the curve Eq. (\ref{critellipse}) only for $\pi/4 < \theta_o < \pi/2$; the continuation of that curve is shown by a dashed red line. Below the solid black line, the dynamics are orientation-preserving; the dynamics also preserve orientation in a small region above this curve for $\theta_0 > \pi/4$, and this region is demarcated by another dashed red line. The period of periodic orbits increases with $E$ and $\theta_o$ until the line $\theta_o=\pi/4$ is approached, and the pattern begins anew for $\theta_o > \pi/4$. When $\pi/\theta_o \in \mathbb{Z}$, there can be exchange-of-interval type orbits which are topologically transitive on a finite set of intervals. The insets give examples in the configuration space for the two phases. }
\label{ellphase}
\end{figure}

Much about the attractor structure for maps with piecewise-negative Schwarzian derivatives has been described in a recent work~\cite{bpp2016}. In this regime we expect strong attractors in microorganism elliptical billiards without orientation reversals, with all values of $q$ belonging to at least one basin of attraction, and attractors primarily of periodic or change-of-interval type. We confirm this expectation and discuss attractors and hyperbolic dynamics in the next section.

\subsection{\label{subsec:ellhyp} Billiard simulation and hyperbolic dynamics}

We choose to determine the values of $(E,\theta_o)$ where the one-dimensional dynamics requires reflection maps $R$ to complete words representing orbits via simulation. We simulate the full microorganism billiards follwing a standard method: we embed the billiard table in the Euclidean plane and calculate the itinerary $\{ q_i \}$ of a trajectory by finding the intersection point of the straight line departing from initial position $q_0$ with angle $\theta_o$ via Newton-Raphson type methods and continuing in this manner to find $q_i$ $\forall i<N$, where $N$ is the maximum number of collisions considered for the orbit with the given initial position. To get an overall picture of the orbit structure, we simulated the billiard on a $30 \times 30$ grid evenly distributed on  table eccentricity $0<E<1$ and outgoing angle $0.1<\theta_o<\pi/2-0.01$, with $N=1000$ and 5-10 different initial positions for each billiard table. The orbit behavior is organized in a phase diagram shown in Fig. \ref{ellphase}. Roughly half of the area of the parameter space describes dynamics which preserve orientation, so that the billiard is governed exactly by Sec. \ref{subsec:ellone}. However, the region of negative Schwarzian derivative only preserves orientation for $\theta_o > \pi/4$ and $E \lesssim 0.96$.

For piecewise negative Schwarzian maps, the maximum number of attractors~\cite{bpp2016} can be as large as $2^{2 \ \mathrm{card}(Sing(F))}$, which for a generic elliptical microorganism billiard can be very large. Interestingly, our simulations never revealed more than one attractor, with a basin of attraction including the entire boundary. The set of periodic attractors has full measure in the parameter space $(E,\theta_o)$ --- however, there are change-of-interval type orbits when $\pi/\theta_o \in \mathbb{Z}$. This is true as well in the orientation-reversing region with negative Schwarzian, and also in a small region of positive Schwarzian derivative for the one-dimensional map with $\theta_o < \pi/4$. Excluding this region, the rest of the orientation-preserving maps with positive Schwarzian derivative have identical structure (apart from $\pi/\theta_o \in \mathbb{Z}$), which we call ``rotator-like'' orbits due to their orbits staying exclusively outside the foci, in analogy with the classical elliptical billiard. These billiards have a caustic which is itself somewhat elliptical in appearance, but rotated with respect to the table boundary. Note that since these tables preserve orientation, their dynamics are equivalently defined by the one-dimensional map and therefore the inverse, which has everywhere negative Schwarzian, also satisfies the conditions given~\cite{bpp2016} to have a finite set of attractors with global basin between them of periodic or change-of-interval type. Yet here we again find only single attractor in the reverse dynamics; all the rotator-like orbits with $\pi/\theta_o \notin \mathbb{Z}$ have periodic orbits as limit sets of the inverse map. 

\begin{figure}
\includegraphics[width=0.5\textwidth]{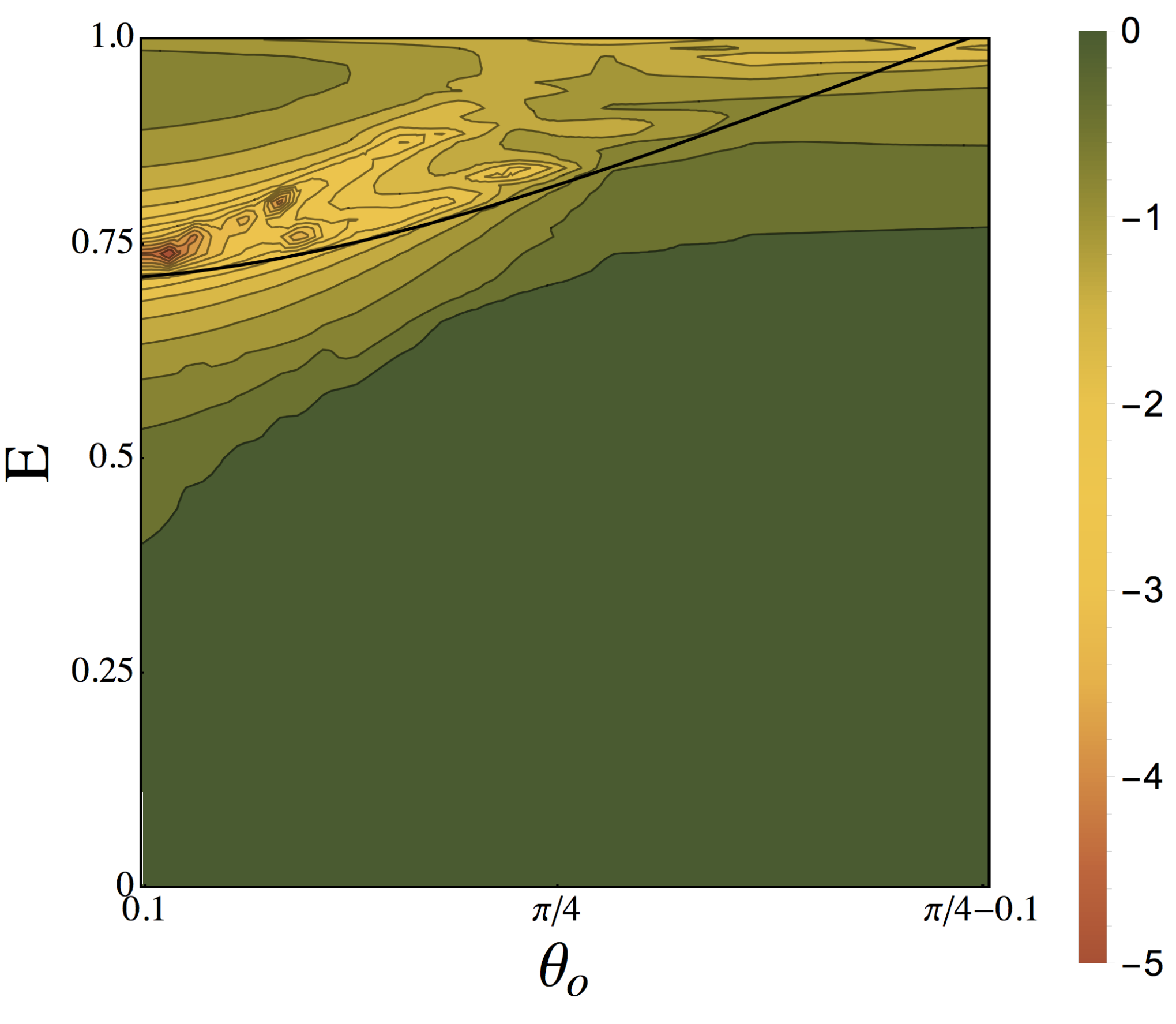}
\caption{(Color online) A surface plot of the Lyapunov exponent for microorganism billiards in the ellipse averaged over three initial conditions. Deep in the rotator-like phase, the Lyapunov exponents are all zero. Nearing the curve Eq. (\ref{critellipse}), the exponent becomes negative; despite the map having positive Schwarzian, the orbits in this regime are attracted to periodic orbits as shown in Fig. \ref{ellphase}. After crossing the curve, the Schwarzian derivative is negative on every component, leading to negative Lyapunov exponent --- the strongest attractors are 4- and 6-periodic orbits which do not preserve orientation. }
\label{lyell}
\end{figure} 


We also calculate the Lyapunov exponent, using a method identical to that used in previous work on microorganism billiards~\cite{wlst2016}. Due to the presence of global attractors for high $E$, we expect the Lyapunov exponent to be negative. This is indeed true, with the results shown in Fig. \ref{lyell}. The curve of the most significance for the Lyapunov exponent seems to be the solid black line in Fig. \ref{ellphase}, which includes some regions of both positive and negative Schwarzian derivative, but only displays periodic orbits. These are the strongest attractors, with 4- and 6-periodic orbits being exceptionally strong. Beneath this curve, the Lyapunov exponents vanish, except for a small region of rotator-like orbits which seem to have slightly non-zero Lyapunov exponents; this could be due to numerical error, or could signal the gradual formation of periodic orbits.  

Lastly, we considered the role of a noisy outgoing angle. Noise-mediated transitions between basins of attraction~\cite{ms1995} seem not to be possible because each table has only one global attractor. To check this fact, we simulated several more initial conditions for each of our parameter points with weak noise, to see if orbits transitioned between multiple attractors. One attractor was seen, with the orbit slightly ``smeared'' by weak noise. If multiple attractors coexisted we did not detect them; since the attractors seem to have global basins, the role of noise in these tables is trivial.

\section{\label{sec:stadium}Bunimovich systems}
\subsection{\label{subsec:stadtheory}Wavefront dynamics}

In this section we consider a class of chaotic billiards where the mechanism for hyperbolicity and positive Lyapunov exponents is that of \textit{defocusing}. This mechanism was first discovered by Bunimovich~\cite{bunim1974a}, and so systems employing the mechanism are often referred to as Bunimovich systems. The most generic requirement for the mechanism to exist in a billiard table is the presence of \textit{absolutely focusing arcs}~\cite{bunim1992}. These systems therefore include an immensely large number of billiard tables; since our results are primarily of a computational nature, we focus on the stadium~\cite{bunim1979} when speaking of hyperbolic dynamics, and then focus on an example of the applicability of billiards theory to practical problems of microbiology by presenting a mushroom-shaped billiard as a trap for microorganisms.

\begin{figure}
\includegraphics[width=0.5\textwidth]{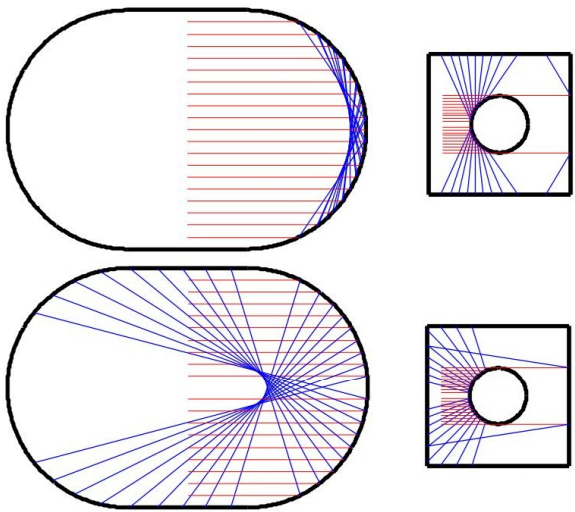}
\caption{(Color online)  Illustration of ray dynamics in the Bunimovich stadium (left) and Sinai unit cell billiard (right, see Sec. \ref{sec:sinai}), for $\theta_o=\pi/3$ (top) and $\theta_o = \pi/20$ (bottom). Rays which would contact the table along the normal have been slightly perturbed to avoid this ill-defined event. The wavefronts in question are locally normal to the ray packets.}
\label{scatter}
\end{figure}

The stadium is constructed using two half-circles of radius $r$ connected by straight lines of length $L$. The family of stadia is therefore defined by a nondimensional parameter $L/r$~\cite{Note2}. In the limit $L/r \rightarrow 0$, we recover microorganism billiards in a circle. For nonzero $L/r$, it is expected that a billiard trajectory will change its orientation with respect to the table during its orbit; as a consequence, we neglect the one-dimensional map given in Eq. \eqref{onedimmap}. As described in Sec. \ref{sec:ellipse}, it would be possible to describe the dynamics using words on the one-dimensional map $F$ and a map $R$ which takes advantage of the four-fold symmetry to effect reversals of orientation in the actual billiard, but we prefer to use the language of wavefronts to provide supporting intuition for our simulations. The wavefront in question is the plane curve which is locally normal to each trajectory in a fan of initially-nearby trajectories, so changes in curvature of a flat wavefront indicate that initially parallel trajectories are either converging or diverging after a collision, see Fig. \ref{scatter}. 

We follow standard conventions~\cite{chernmakar}, according to which the wavefront curvature $\mathcal{B}$ is defined to be zero for a flat wave, positive for a dispersing/expanding wave, and negative for a focusing/contracting wave; at a focusing point of the wave, $\mathcal{B}=\infty$. We denote the curvature of a front immediately before a collision as $\mathcal{B}^-$ and immediately after as $\mathcal{B}^+$. The two are related by an aspecular variant of the mirror equation~\cite{mps2009}:
\begin{equation}
\mathcal{B}^+=\frac{K}{\cos \theta_o}. \label{mirroreqn}
\end{equation}
Note that the specular mirror equation reads $\mathcal{B}^+ = \mathcal{B}^- + \frac{2 K}{\cos \theta_o}$; in particular, our outgoing curvature does not depend at all on the incoming curvature. During the subsequent free-flight, the curvature evolves according to 
\begin{equation}
\mathcal{B}_t = \frac{1}{t+1/\mathcal{B}^+}. \label{freeflight}
\end{equation}
These two expressions can be combined in a continued fraction formula to give the curvature of the front evolving at any time during the billiard dynamics. These dynamics will be very different from the specular limit due to the absence of $\mathcal{B}^-$ in Eq. (\ref{mirroreqn}); in particular, a front ``forgets'' its past during any collision and adopts a new curvature which is sensitive to $\theta_o$. The aforementioned defocusing mechanism is based on an analysis of Eq. (\ref{freeflight}) which reveals that a focusing wave front actually expands between two collisions taking place at a distance $\tau$ apart so long as 
\begin{equation}
\tau \mathcal{B}^+_0<-2, \label{defocus}
\end{equation}
 where $\mathcal{B}^+_0$ is the curvature emanating from the first of the two collisions.

The general theory of how hyperbolicity and chaos arise in Bunimovich billiards reveals which type of collision events lead to positive Lyapunov exponents, so we will analyze some basic situations using Eq. (\ref{mirroreqn}) and Eq. (\ref{freeflight}) to determine when Eq. (\ref{defocus}) is satisfied. We first divide the types of collisions into three possible classes: \textbf{i)}, the microorganism leaves a point on a circular arc and contacts another point on the same circular arc, \textbf{ii)} the microorganism leaves a point on one flat component and contacts a point on another flat component, \textbf{iii)} the microorganism leaves a point on one circular arc and contacts a point on one flat component, or \textbf{iv)} the microorganism leaves one circular arc and contacts another. In the specular theory, (i),(ii) and (iii) are often called ``nonessential collisions'', as one can show that the curvature of wavefronts remains unchanged by these collisions~\cite{chernmakar}. In microorganism billiards, however, (ii) and (iii) immediately ``reset'' the curvature of the wavefront to zero, violating (\ref{defocus}). If the radius of a circular cap is $r$, then an event of type (i) or (iv) is only defocusing if
\begin{equation}
-\frac{\tau}{r \cos \theta_o} <-2,
\end{equation}
so generically speaking type (iv) events are more likely to lead to defocusing. This leads to an interesting competition between the microorganism outgoing angle and the geometry of the table. For the stadium, having line sections much longer than the radius of the caps will increase the likelihood of defocusing type (iv) effects, but these are also increasingly often ``reset'' by a type (ii) or type (iii) event. We therefore expect shorter line segments to lead to more chaotic dynamics. However, decreasing the length too much will lead to focusing events as $\tau$ decreases; this is especially deleterious if $\theta_o > \pi/4$. In addition, motivated both by our results in Sec. \ref{sec:ellipse} on microorganism billiards in the ellipse and the presence of  marginally unstable periodic orbits (or MUPO's) in the specular Bunimovich tables, we expect that the hyperbolic dynamics will be complicated by the presence of attractive periodic orbits --- therefore the Lyapunov exponent is expected to be highly sensitive to table geometry, outgoing angle, and initial conditions. We do not attempt to demonstrate uniform hyperbolicity for these tables or for the tables in Sec. \ref{sec:sinai}, as it is obvious in the case of the stadium that uniform hyperbolicity does not exist for the vast majority of geometries; we discuss uniform hyperbolicity for Sinai tables very briefly, as it is obvious when it will be preserved or destroyed by the aspecular reflection law. 

\begin{figure}
\includegraphics[width=0.5\textwidth]{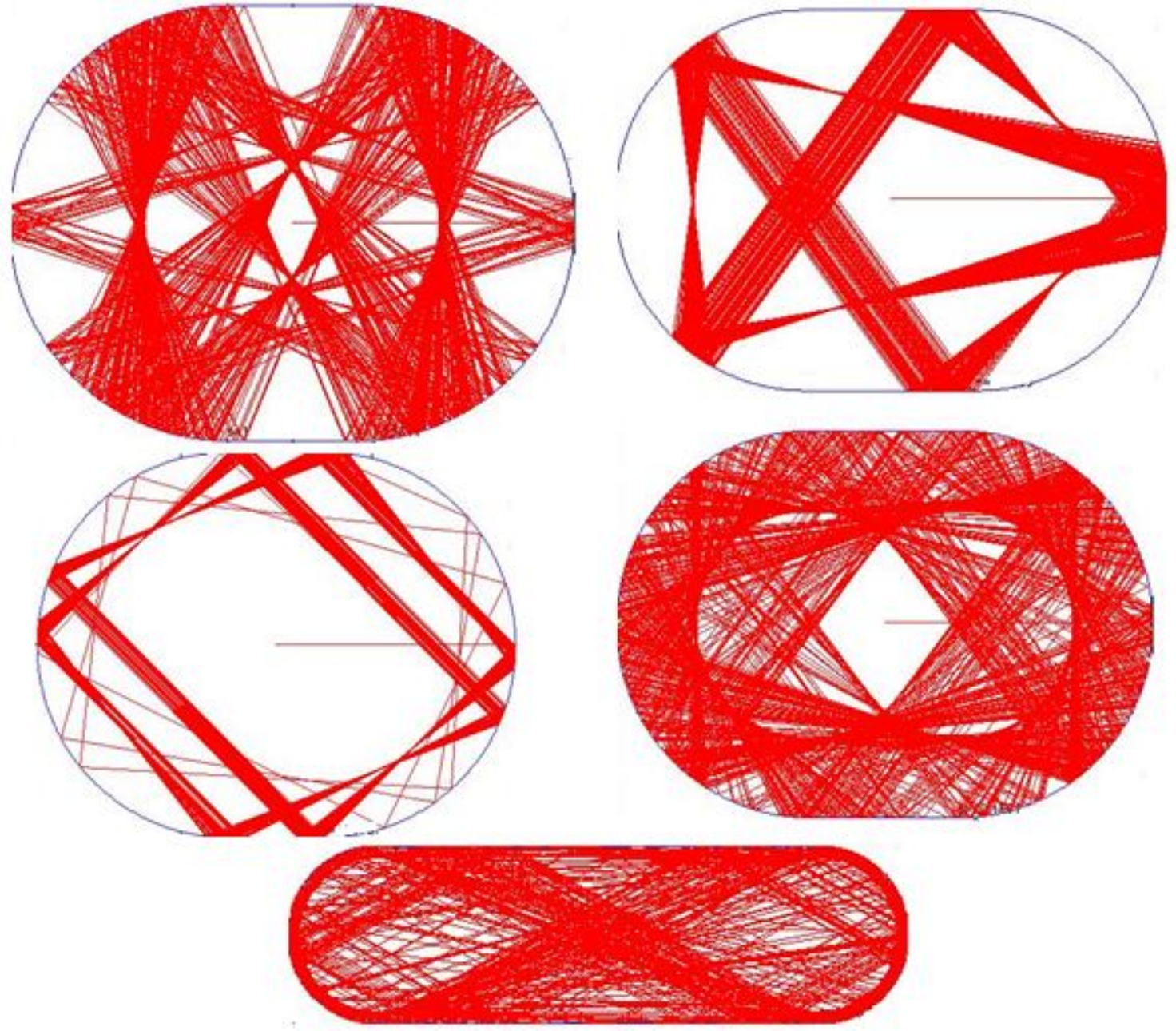}
\caption{(Color online)  A sampling of configuration spaces for microorganism billiards in stadia.}
\label{stadcaps}
\end{figure}

\subsection{\label{subsec:stadsim}Billiard simulation and hyperbolic dynamics}


\begin{figure}
\includegraphics[width=0.5\textwidth]{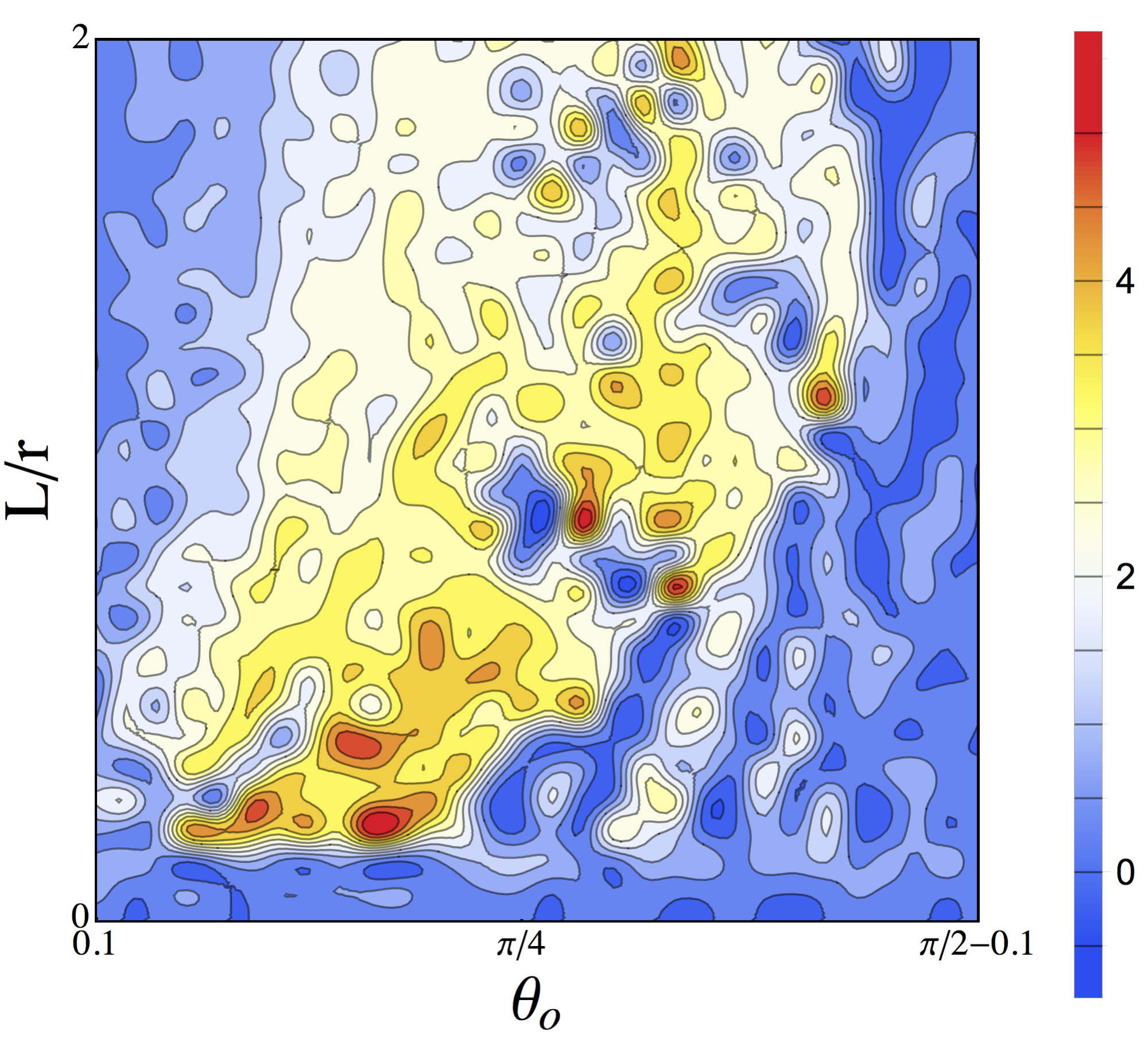}
\caption{(Color online) Contour plot of the Lyapunov exponent in the stadium averaged over three initial conditions chosen from the uniform distribution for various geometries and outgoing angles. They are generically much higher (approximately 5-10 times higher) than in the specular case~\cite{dp1995} due to the removal of nonessential collisons. However, much unlike the specular case, pockets with negative Lyapunov exponent can be found, especially for $\theta_o > \pi/4$ --- these correspond to weakly attracting periodic orbits (marginally unstable periodic orbits, or MUPO's) comprised only of nonessential collisions. The introduction of noise destabilizes these orbits, resulting in a Lyapunov exponent landscape which is higher but also much more homogeneous.}
\label{lystad}
\end{figure} 

We studied the orbit types and hyperbolic dynamics of microorganism billiards in the Bunimovich stadium by simulating~\footnote{The systems of interest are hyperbolic, so we assume that the shadowing lemma applies to ensure the applicability of our simulations.} the dynamics in a $30 \times 30$ grid distributed evenly on the table geometry $L/r \in [0,2]$ and the outgoing angle $\theta_o \in [0.1,\pi/2-0.1]$ for several initial conditions. Again, the bounds on $\theta_o$ have been selected to avoid ill-defined events such as orbits tracing the boundary of the table ($\theta_o=\pi/2$) or the slap map ($\theta_o=0$), which for the stadium simply converges to the two-periodic attractors representing only-flat and only-cap collisions. Unlike the relatively simple orbit structure of microorganism ellipse billiards (Sec. \ref{sec:ellipse}) or the generically-chaotic orbit structure of Sinai billiards (Sec. \ref{sec:sinai}), the stadium showed an incredibly rich array of orbits. We estimate that there are between 15 and 25 distinct orbit types, including periodic orbits with a range of periods, some resembling ``cuts'' of the specular stadium orbit --- by this we mean any coherent segment of links which could occur in a specular stadium trajectory that is compatible with our reflection law, repeated periodically. These latter structures are easily recognized by plotting the phase space trajectory for several initial conditions and comparing with the phase space of the specular case. We show some representative examples of configuration spaces in Fig. \ref{stadcaps}. Other orbits were less easily classified, representing an interaction between the weakly-attracting periodic orbits and the defocusing trajectories. This competition is reflected in Fig. \ref{lystad}, where we plot the Lyapunov exponent averaged over three initial conditions chosen uniformly from Lebesgue measure on the table domain and initial orientation. For tables nearer the boundary of the figure (representing more extreme geometry/outgoing angle combinations), the Lyapunov exponent is lower than the specular one or even negative; for more generic situations, the Lyapunov exponent can be as much as an order of magnitude higher than the specular case due to the increase in defocusing links (equivalently, the reduction of nonessential collisions). We refer to the attractive periodic orbits as ``weakly attracting'' because for even very low thresholds of noise, such as the uniform distribution $\theta_o/|\theta_o| \in [0.9 , 1.1]$, the Lyapunov exponent landscape homogenizes at a high positive value and no periodic structures can be found in the orbits. This is a strong contrast to the results of Sec. \ref{sec:ellipse}, where noise simply tended to ``smear'' the globally attracting periodic orbits.

\subsection{\label{subsec:stadtrapfil} Design for a microorganism trap}

\begin{figure}
\includegraphics[width=0.3\textwidth]{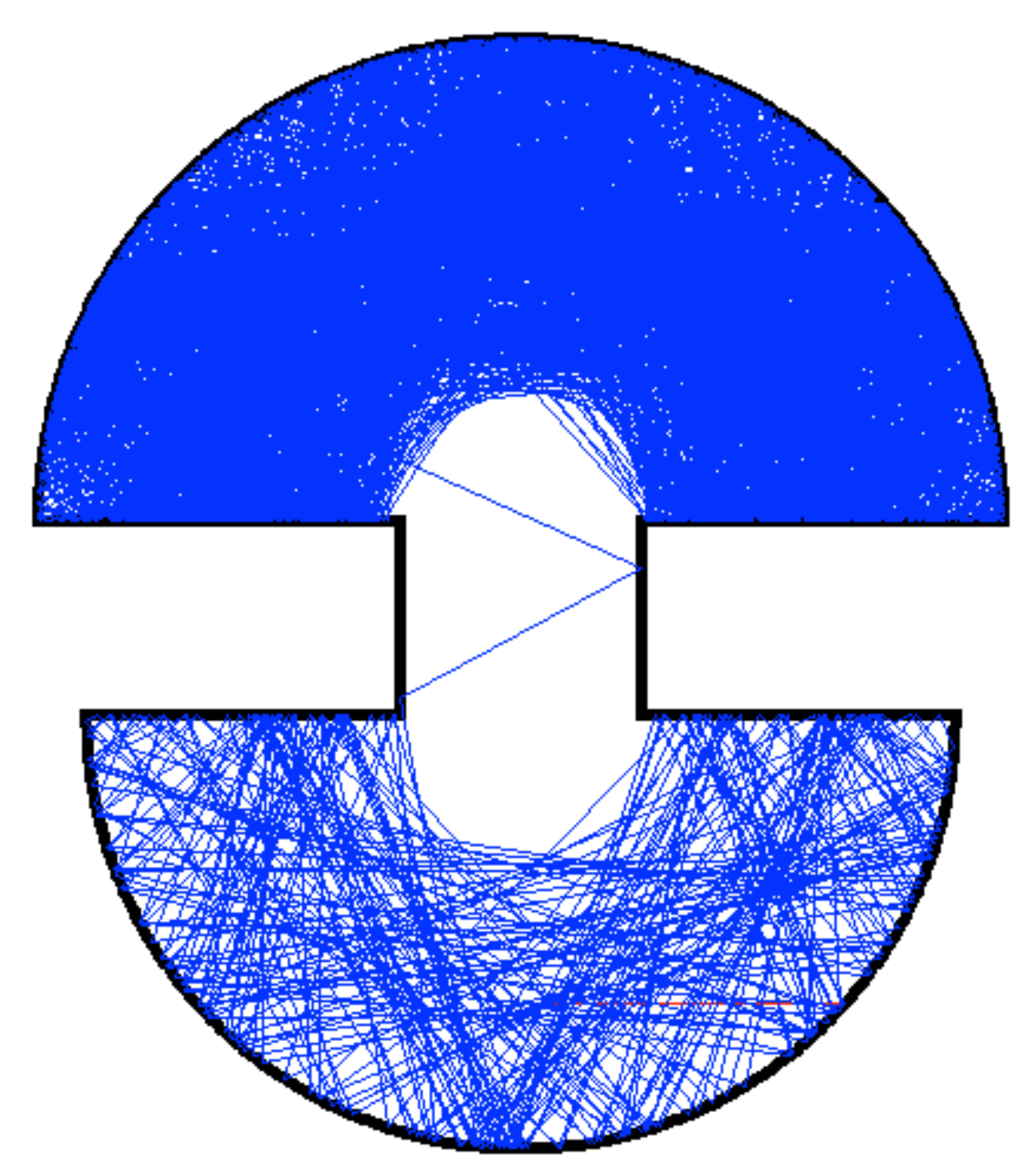}
\caption{(Color online) Microorganism billiards in the double-mushroom; the top lobe has a radius which is $10\%$ larger than that of the bottom lobe. The width of the stem is $20\%$ the radius of the top lobe and the length of the stem is $50\%$ the radius of the top lobe. Here, $\theta_o = \pi/6 + \theta_n$, where $\theta_n$ is uniformly distributed on $(-\pi/12,\pi/12)$. The microorganism begins in the bottom lobe and migrates to the top lobe after a few hundred collisions, where it remains thereafter --- this calculation is continued for $N=3000$ collisions. The initial condition is shown as a red line. If the noise is Gaussian distributed rather than uniformly distributed, the ``trap'' is not permanent; an event from the tail of the distribution will eventually propel the microorganism back into the bottom lobe.}
\label{shroomtrap}
\end{figure} 

Because flat walls and absolutely focusing arcs can be combined in an infinite number of ways, the Bunimovich class of billiards gives the table designer a large amount of control over the properties of the phase space. This is an extremely useful tool in the context of designing a trap for one or more microorganisms, which has been the subject of much recent work~\cite{bjmvdvsetal2,smbl2015,jkp2016}. A prime example of a useful billiard table for constructing traps is a mushroom-shaped table~\cite{bunim2001}, which has a phase space which is divided into an ordered/integrable component comprised of periodic orbits trapped inside the spherical cap, and a chaotic component with orbits that include collisions with the flat walls of the stem and the spherical cap \footnote{Note that depending on definitions the mushroom may not qualify as a true Bunimovich billiard; the system is usually equipped with a condition that the completion of any circular arc must lie entirely within the table boundary}. The boundary between the two regions is usually of a fractal nature, and also `sticky'~\cite{amk2005,dettmann2016}, meaning that orbits originating in the chaotic region can spend large amounts of time at the border with the integrable region due to the existence of marginally unstable periodic orbits, or MUPO's. Motivated by this phase space structure, as well as previous results on the role of noise in trapping events~\cite{ae2010}, we designed a passive, tunable trap for microorganisms based on the mushroom shape. Ours is an asymmetric double-mushroom, pictured in Fig. \ref{shroomtrap}. The mechanism of the trapping depends on both the asymmetry of the table and noise in the outgoing angle of the microorganism. In the limit of zero noise, the behavior of the microorganism is entirely determined by the radii of the two lobes, the outgoing angle and the initial condition: the latter two set a fixed angular momentum $\mathcal{L}$ (defined as the distance of closest approach to the focus of the arc). If $\mathcal{L}$ is larger than the radius of the lobe in which the microorganism originates, the orbit will be periodic and contained in that lobe; if $\mathcal{L}$ is smaller, the microorganism will eventually migrate to the other lobe, where the process is repeated. This description is only approximate, ignoring several fine details~\cite{dettmann2016}, but serves to illustrate the idea that lobe asymmetry can filter microorganisms from an escapable lobe to an inescapable lobe. 

The inclusion of noise and its distribution are important to the design of this trap --- not only do they simplify the picture by removing MUPO's, they allow for rather precise tunability in first-return properties when desirable, or suggest lobe radii for different purposes. In our example, we consider the noise to be uniformly distibuted; this leads to a rather simple design for a trap, for uniform noise in the outgoing angle leads to a bounded distribution of angular momenta; one need simply choose one lobe to have a radius outside of this bound to trap a microorganism for all time. This is the situation pictured in Fig. \ref{shroomtrap}. If the noise is instead Gaussian distributed, events from the tail of the distribution will cause the microorganism to return after many collisions. We leave the precise details of first-return times and their dependency on $\theta_o$ and table geometry to future work. This division of phase space is a property of a robust class of billiard tables, and we believe that it can be very useful in the design of microfluidic devices for various purposes involving mixing, filtering, and trapping of microorganisms based only on table geometry and outgoing angle.

\section{\label{sec:sinai}Sinai Systems}
\subsection{\label{subsec:sinaitheory}Unfoldings and wavefront dynamics}

Our last system of interest is the chaotic billiard system discovered by Sinai~\cite{sinai1963,sinai1970}. The model has deep connections to the fundamentals of statistical mechanics~\cite{chernmakar}. The tables are in some sense ``opposite'' to the Bunimovich tables, in that they are comprised only of dispersing walls (those having positive curvature $K$). This leads to a wealth of tables inhabiting different types of spaces, of which we focus on only one example. We briefly discuss the tools of wavefronts and unfoldings before presenting the results of our simulations, comparing the Lyapunov exponents obtained to other works in the literature~\cite{dp1995,dahl1997}.

\begin{figure}
\includegraphics[width=0.5\textwidth]{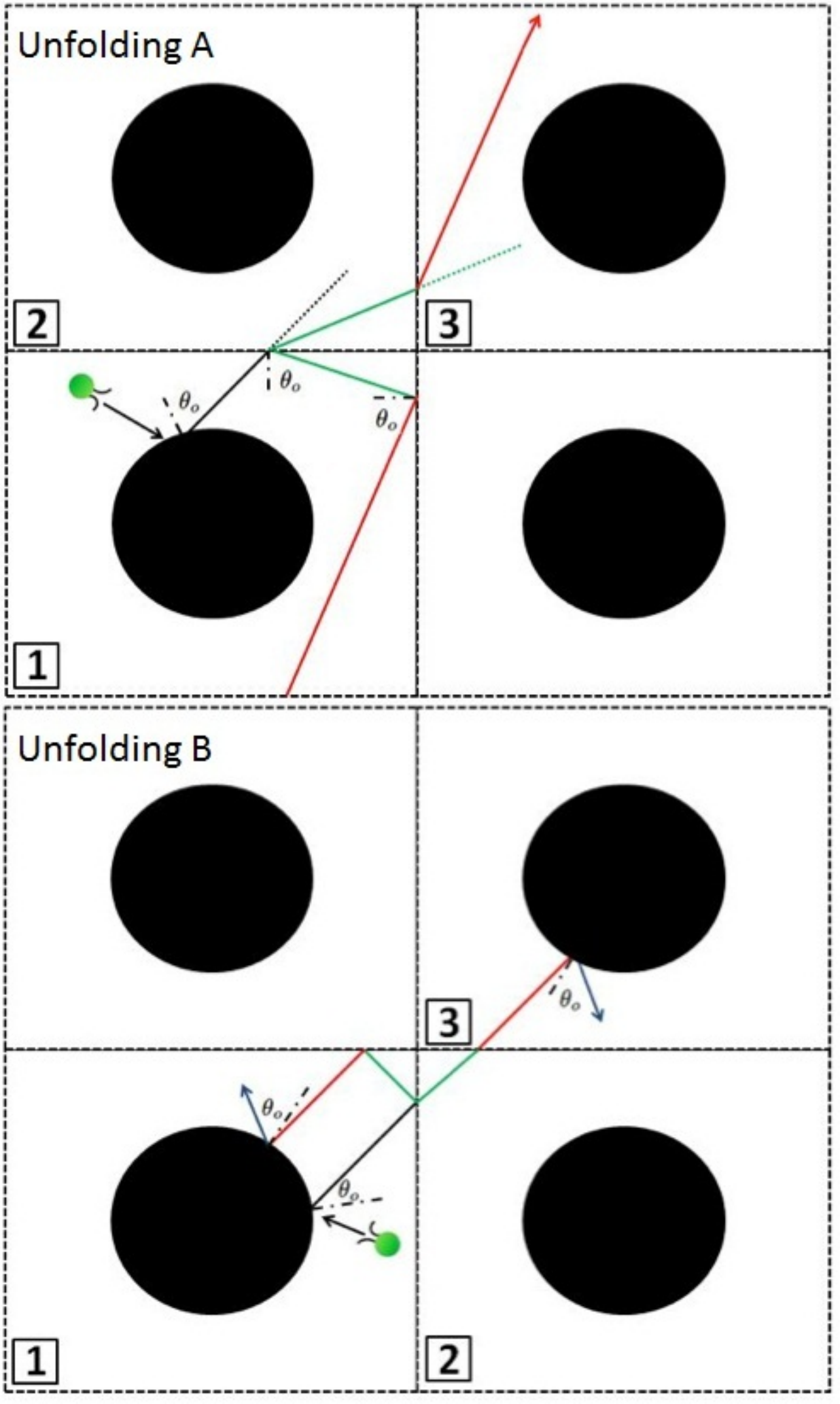}
\caption{(Color online) Two possible unfoldings for the microorganism Sinai billiard. The unfolding A naively applies the aspecular reflection law on all boundaries, and may represent a microorganism responding to external stimuli or `tumbling' type events. The unfolding B has no such events; to preserve the direction of free-flight in the lattice, specular reflections are applied on the walls of the unit cell. Numbered labels of cells indicate the order of re-folding to obtain the trajectory in the unit cell.}
\label{unfoldings}
\end{figure} 


\begin{figure*}[t]
\includegraphics[width=\textwidth]{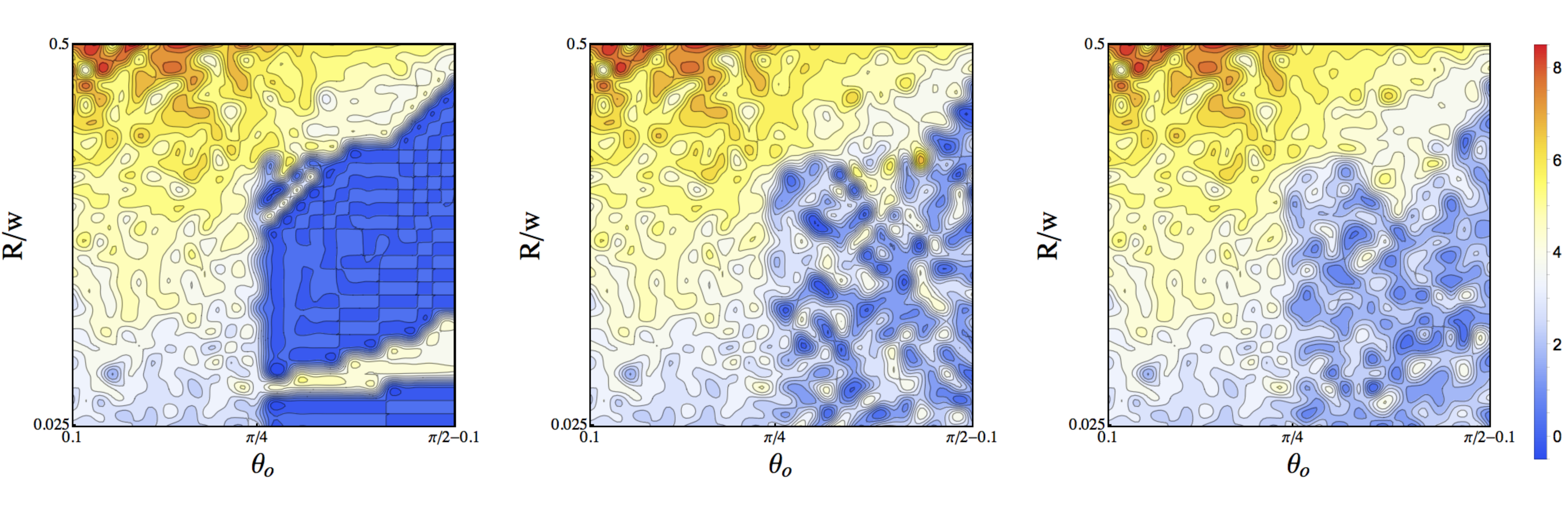}
\caption{(Color online) Lyapunov exponents for the A-type unfolding for various geometries and outgoing angles. Moving from the leftmost frame to the rightmost increases the number of initial conditions being averaged over --- three (left), six (center), and nine (right). The regions of negative Lyapunov exponent, corresponding to the attracting periodic orbits in the square boundary of the primitive cell~\cite{wlst2016} which contain the scatterer, do not completely break up even when averaging over many initial conditions chosen uniformly. Away from this region the Lyapunov exponents are similar to those measured in other works~\cite{dp1995,dahl1997}; they are somewhat lower for large $R/w$ and higher for small $R/w$ in comparison with the specular case, indicating that the microorganism can more easily ``find'' even a very small scattering disk.}
\label{sinaitriptych}
\end{figure*}

\begin{figure}
\includegraphics[width=0.5\textwidth]{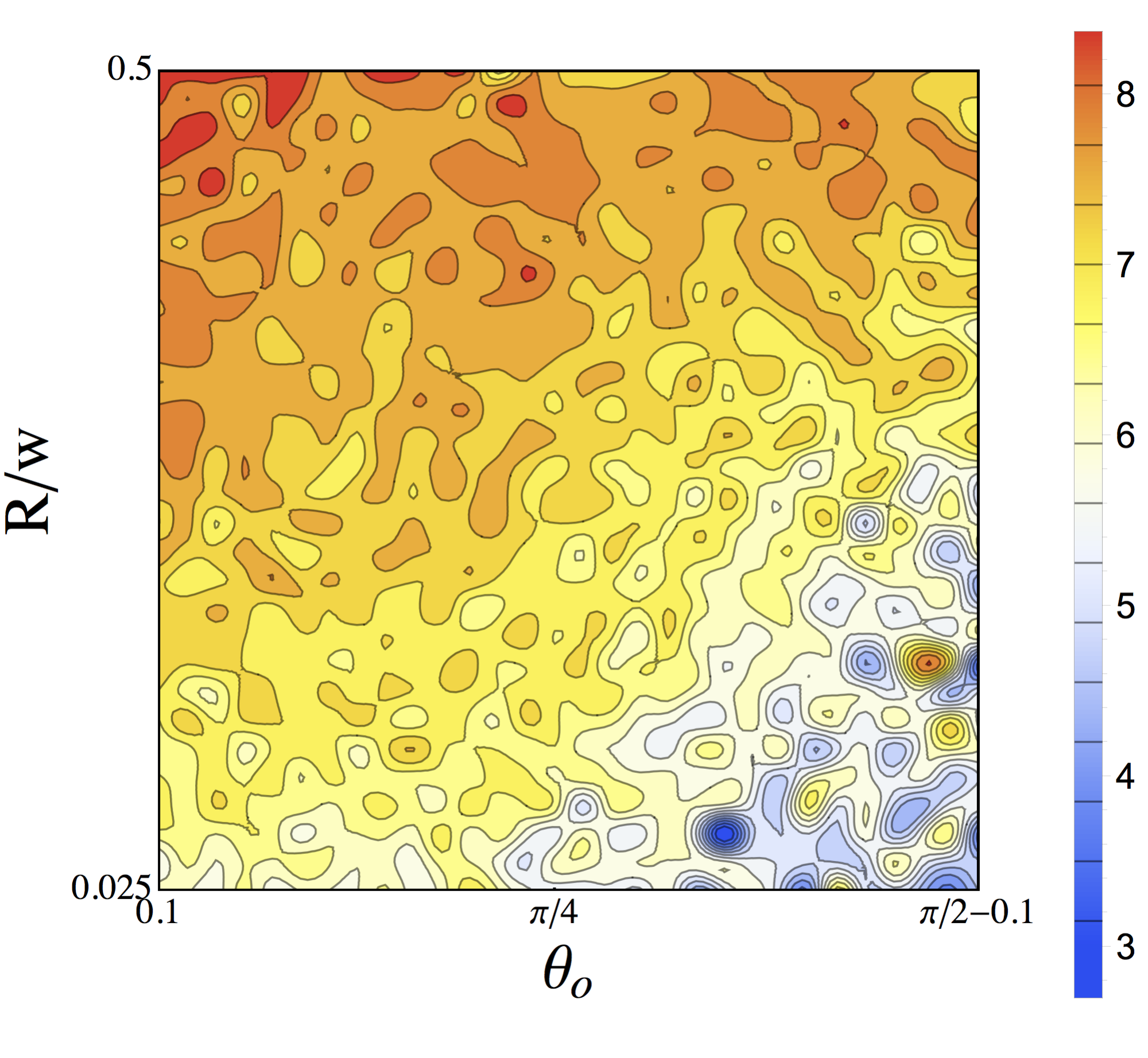}
\caption{(Color online) Lyapunov exponents for the B-type unfolding for various geometries and outgoing angles, averaged over three initial conditions. Equipping the walls of the primitive cell with a specular reflection law leads to higher Lyapunov exponents (more scattering events) than the A-type unfolding. }
\label{lysinb}
\end{figure} 

Our analysis begins with Eq. (\ref{mirroreqn}). Combining this with Eq. (\ref{freeflight}) reveals that wavefront curvature decreases hyperbolically in free-flight, asymptoting to zero, and jumps to a fixed positive number at collisions with dispersing walls or to zero at collisions with a flat wall, tracing an erratic saw-toothed curve in time. As an example, we consider single dispersing disk of radius $R$ centered in a square cell of length $w$~\cite{Note2}. By creating copies of this unit cell and reflecting collisions from the flat walls across the unit cell, we arrive at an \textit{unfolding} of the unit cell in which the billiard particle now navigates a square lattice of scatterers of radius $R$ and lattice spacing $w$. This geometry therefore captures the dynamics of two billiard tables at once. If we naively play microorganism billiards, however, the unfolded problem illustrates a rather strange feature, that the microorganism at some points in free-flight suddenly ``decides'' to reorient itself in response to the imaginary line representing the wall of the unit cell. Because of its deterministic character this is unlike a run-and-tumble event; we call this billiard the ``drunken swimmer'', or Unfolding A (see Fig. \ref{unfoldings}). As a representative of a relevant physical system, the unit cell dynamics of Unfolding A are likely more realistic than the lattice dynamics. We therefore study a second pair of unit cell-lattice systems, one which has aspecular reflections on scatters and specular reflections on flat walls, which we call Unfolding B. This latter system is more realistic for microorganisms navigating lattices of scatterers, which has been a subject of previous inquiry~\cite{wrnr2008,gkca2007}. We further note that this billiard system is of physical interest due to its previous use to calculate rheological quantities~\cite{bs1996}, which in our case are expected to be highly dependent on the outgoing angle --- we leave such simulation work to future endeavors. 

It is worth noting immediately that Unfolding B has wave dynamics very similar to the specular Sinai tables; it has a curvature which stays bounded away from zero, $\mathcal{B}_{min} = \frac{1}{\tau_{max}+1/\mathcal{R}}$, where $\mathcal{R} = \frac{K}{\cos \theta_o}$, and of course a maximum $\mathcal{B}_{\max}=\mathcal{R}$. Exponential growth of the length of the wavefront and uniform hyperbolicity follow generally, where the exponent in the growth of wavefront length is sensitive to $\theta_o$. Unfolding A is not uniformly hyperbolic due to the ``resetting'' of the curvature. Moreover, the existence of strongly attracting periodic orbits in the square microorganism billiard~\cite{wlst2016} implies the existence of attractors which \textit{avoid} the scatterer for certain values of $R/w$ and certain initial conditions; depending on the geometry and the strength of the attractor, this can lead to entirely polygonal dynamics, and in the lattice problem these correspond to ``drunken'' swimmers which are able to successfully navigate the lattice of scatterers without collisions. Such events are expected to dramatically decrease the Lyapunov exponent measured in these tables. Because our systems are already highly chaotic, we do not consider noise; it will have a strong effect on the periodic attractors in Unfolding A, and we expect the full implications of noise to be represented by results in other work~\cite{shsl2009}.

\subsection{\label{subsec:sinaisim}Billiard simulation and hyperbolic dynamics}

As in the previous sections, we simulated the microorganism billiard dynamics on a $30 \times 30$ grid, here distributed evenly on $R/w \in [0.025, 0.5]$ and $\theta_o \in [0.1, \pi/2-0.1]$, where $R$ is the radius of the scatterer and $w$ the width of the cell; the upper bound of our parameter $R/w$ therefore corresponds to the limit of an absolutely finite horizon, where the particle is trapped in one quadrant of the unit cell or between four abutting scatterers in the unfolded lattice. We do not show any orbits; in both unfoldings, the trajectories from several initial conditions in both configuration and phase spaces are visually identical to those in the specular case, and seem to be statistically identical, though we cannot say so precisely without well-defined invariant measures. Note that with our reflection law links between two parallel flat surfaces happen to reflect specularly, so both Unfolding A and Unfolding B alternate between specular and aspecular reflections, though Unfolding A will have more aspecular events. 

We plot the corresponding Lyapunov exponents for Unfolding A and Unfolding B in Fig. \ref{sinaitriptych} and Fig. \ref{lysinb}, respectively. In both cases the asymptotic dependence on the geometry $R/w$ is identical to that known in the specular case~\cite{dp1995,dahl1997}, but the actual value of the exponent is either larger or smaller than in the specular case depending on outgoing angle. For Unfolding A there is the influence of strongly attracting periodic orbits discussed in previous work~\cite{wlst2016} which do not contact the scatterer. These seem to be rather pervasive; we plot the Lyapunov exponent averaged over an increasing number of initial conditions and still find regions of negative Lyapunov exponent. Unlike similar events in Sec. \ref{sec:stadium}, these are not so easily removed by noise; the range of noise necessary to make all Lyapunov exponents positive, when drawn from a uniform distribution, is proportional to $\frac{w^2 \theta_o}{R^2}$. Unfolding B behaves much more like the specular Sinai billiard, and except for a few instances its Lyapunov exponent landscape is on the same order of magnitude as the specular case.

\section{\label{sec:summary}Summary}

Microorganism billiards, in which the outgoing angle is a constant or random variable, provide an example of a physically relevant non-conservative billiard system. The system is both simple and complicated --- in many table geometries and for many outgoing angles $\theta_o$, the billiard dynamics will reduce to a single one-dimensional map. We have seen examples where this map has negative Schwarzian derivative everywhere, a condition for which much theoretical work in one-dimensional maps has been accomplished. On the other hand, there is no simple invariant measure, so that many of the standard tools and theorems are not applicable, complicating the cases when the dynamics are words on two generators, the one-dimensional map and rotations which account for orientation reversals due to the microorganisms preserving their forward (tangential) momentum. The system is of interest not only as a novel billiard system, but also because several interest results from billiard theory can be made useful in the study of swimming microorganisms. Bunimovich tables and more generally tables with divided phase space can be utilized to design robust classes of traps, providing essentially infinite tunability to first-passage times and methods of combining and filtering different strains of microorganisms. Sinai tables provide insight into scattering by a lattice of obstacles, and will form a fundamental template for studying and optimizing scattering by differently shaped obstacles in future work. We hope that our work will be of some value to both the study of billiards and the physics of swimming microorganisms.

 \bibliography{newrefs}

\end{document}